\documentclass[prb,twocolumn,showpacs,floatfix]{revtex4}
\usepackage{amsmath}
\usepackage{graphicx}
\usepackage{epsfig}
\usepackage{dcolumn}% Align table columns on decimal point
\usepackage{bm}% bold math
\newcommand{\be}{\begin{eqnarray}}
\newcommand{\ee}{\end{eqnarray}}
\newcommand{\bfq}{\bf{ q}}

\begin{document}

\title{Finite Drude weight for 1D low temperature conductors}

\author{Dariush  Heidarian  and Sandro Sorella}

\affiliation{Istituto Nazionale di Fisica della Materia (INFM)-Democritos,
National Simulation Centre, and Scuola Internazionale Superiore di Studi
Avanzati (SISSA), I-34014 Trieste, Italy}

\begin{abstract}
We apply  well established finite temperature 
Quantum Monte Carlo techniques to   
one dimensional Bose systems with soft and hardcore constraint, 
as well as to spinless fermion systems. 
We give clear and robust  numerical evidence that, as expected,     
no superfluid density for Bosons or Meissner fraction for fermions.   
is possible at {\em any} non zero temperature in one  dimensional 
interacting Bose or fermi lattice models, whereas 
a finite Drude weight is generally observed in gapless systems, 
in partial disagreement to previous expectations. 
\end{abstract}
%\keywords      {Bose metals, superfluididty, cold atoms}

\maketitle

%%%%%%%%%%%%%%%%%%%%%%%%%%%%%%%%%%%%%%%%%%%%
%% MAINMATTER
%%%%%%%%%%%%%%%%%%%%%%%%%%%%%%%%%%%%%%%%%%%%

\section{Introduction}
In the last decades there have been a lot of numerical and theoretical 
works  to understand the role of strong correlation 
in    lattice 
model Hamiltonians.\cite{ceperley,scalettar,dmrgbose,troyer,sandvik,fisher,cazalilla} 
Recently this issue has acquired an increasing attention and remarkable 
importance, due to 
the recent advances  in the realization of optical lattices. In these 
experiments  ultracold atoms behave as boson particles  
trapped on particular lattice sites,  whereas  the interaction and the 
hopping parameters can be tuned continuously. 
This important achievement  has opened the   possibility to verify 
 directly the crucial role played by the electron correlation in 
% by validating or falsifying long standing   theoretical 
%predictions on strongly correlated 
%model Hamiltonians, 
very important model Hamiltonians defined on a lattice. 
An important example is the realization of a Mott insulating state in 
a system with strong on site repulsion\cite{greiner2d,greiner3d}. 
Moreover quite recently the 
possibility to include  the Fermi statistics 
in optical lattices 
appears very promising  and interesting.\cite{boh}

In 1D spinless fermion systems are equivalent to interacting Bose systems 
with hard-core constraint and are described by the same low energy theory
-the Luttinger liquid theory-.
Indeed this theory holds  also for 
soft-core bosons, as shown in  Ref.(\onlinecite{cazalilla}). 
Therefore, as far as the 
transport properties are concerned one should expect the same 
behavior both for fermions and bosons.
On the other  hand for lattice models, even in absence of disorder, 
the current does  not
commute with the Hamiltonian, implying its  possible decay  
at finite temperature due to the backscattering processes\cite{andrei}.
In this case the dynamical 
 current-current correlation function also decays in time,
leading to a current Fourier transform without  $\delta $ function
at zero energy, namely without a finite  Drude weight 
within the linear response theory. 

Until few decades ago the absence of the Drude weight  was the 
expected behavior of all interacting  metals  in lattice models or in real
solids at finite temperature.  
However a quite clear numerical evidence has 
been reported in Ref.\onlinecite{zotos}   that current should not decay 
in integrable 1D models, namely for Hamiltonians   that 
can be solved by Bethe ansatz techniques in 1D. These models  
essentially  
possess some hidden conservation law, that was conjectured to 
forbid the current decay process.\cite{zotos,zotos2}  
Later several groups have reproduced this 
surprising effect\cite{poilblanc,hanke}, with a noticeable exception that 
a finite Drude weight at finite temperature was found 
also for  non-integrable models.\cite{hanke}  
On the other hand, from purely theoretical grounds  this  issue is not  
settled yet: 
in Ref.\onlinecite{andrei}
it was argued that backscattering processes can be effective also at 
finite temperature and in 1D non integrable models,
 whereas in Ref.\onlinecite{kawakami}, it was proposed 
that also some particular 
 non integrable model could provide a conserved current.

In this work we propose that the general behavior of 1D gapless 
systems  is eventually characterized by a {\em finite} Drude weight at finite temperature, and we have found no exception in the models that we have studied.
This conclusion is based on a careful and systematic numerical work 
on fairly generic one dimensional Bose and Fermi systems, that {\em all} show 
the same behavior, even though strong finite size effects are observed 
in the non integrable cases.

In the following  we investigate the behavior of the Drude weight in 1D
systems in the thermodynamic limit and finite temperature. 

$Model~and~Method:$ We have studied hardcore and softcore bosons in a 1D
lattice with periodic boundary conditions. The Hamiltonian studied  reads, 
\be
H=\sum_{i}\biggl(-t(a^\dagger_ia_{i+1}+h.c.)+
\frac{U}{2}n_i(n_i-1)\nonumber\\
+Vn_in_{i+1}+Wn_in_{i+2}-\mu n_i\biggr)
\ee
The sum is over all lattice sites $i$, $a_i^\dagger/a_i$ is the boson
creation/annihilation operator at site $i$, henceforth 
$n_i$ is the particle number at site $i$ and$\mu$ is the chemical potential. 
$t$ is the hopping amplitude
which is set to one, $U$ is the on-site
repulsion, whereas  $V$ and $W$ are the nearest and the 
next-nearest neighbor interactions, respectively. 
For hardcore bosons in the $U\rightarrow\infty$ limit 
the  Hamiltonian can
be mapped onto an $S=1/2$ spin system with $S^z_i=n_i-1/2$ and
$S_i^+=a_i^\dagger$.  
In this work  we present our
results for the half filled case of hardcore and softcore models. 
Most of our 
  results have been obtained by  Quantum Monte Carlo (QMC),  using  the 
stochastic series expansion (SSE)\cite{sandvik, sandvik2} with
the directed loop update\cite{syljuasen}.

Superfluid density $\rho_s$ (or spin stiffness in the equivalent spin model), 
is defined as the second derivative of the
free energy with
respect to a twist in the boundary conditions. 
In order to compute this quantity by QMC, it is convenient to 
apply  linear  response theory, relating this quantity to the 
current current response function $\Lambda({\bfq},i\omega_n)=
\int_0^\beta d\tau\exp(i\omega_n\tau)\langle
J({\bfq},\tau)J(-{\bfq},0)\rangle/N$, where $J$ is the current operator and
$\omega_n$ is Matsubara frequency. 
Then  the following expression for the superfluid density is obtained:
\be 
\rho_s=\langle -K\rangle -\Lambda(q=0;i\omega_n=0)=\frac{\langle
W^2\rangle}{\beta}
\ee
where $\langle K\rangle$ is the average kinetic energy per site,
$\omega_n=2\pi n/\beta$ are the  Matsubara frequencies  and $W$ 
is the winding number.  Similarly  
the Drude weight is obtained with the same expression  
but  with a different order in the  limit $\omega\rightarrow0$ and 
$q\rightarrow0$, namely\cite{scalapino,hanke,note} 
\be
D=\langle-K\rangle-\mbox{Re}\Lambda(q=0,\omega\rightarrow 0).
\ee 
In SSE one can obtain $\Lambda$ very accurately  in terms of Matsubara
frequencies. Therefore analytic continuation
of the data is required. In order to avoid difficulties of extrapolation to 
$i\omega_n\rightarrow0$ at large temperatures, we have worked 
at relatively low temperatures ($\beta\geq10$).

In principle,  due to the different order of limits,
the  Drude weight and the superfluid density may be different 
when the following quantity remains finite in the thermodynamic limit\cite{hanke}: 
$D-\rho_s=\sum_{E_n=E_m}\beta\exp(-\beta 
E_n)|\langle\psi_n|J|\psi_m\rangle|^2/L$, 
where, $J$ is the
current operator, while 
 $E_n$ and 
$|\psi_n\rangle$ are the $n^{th}$ eigenvalue and 
eigenstate of the many body system, respectively. 

The current operator can be 
 written as $J(q=0)=i\sum_b(H_{b}^+-H_{b}^-)$ where
$H_{b}^+=ta_{l}^\dagger a_{l+1}$ and $b$
is the bond index, corresponding to the site index $l$. 
The ensemble average of product of two local operators
$H_{b_1}^{\sigma_1}$ and $H_{b_2}^{\sigma_2}(\tau)$ is:
\be
\langle H_{b_2}^{\sigma_2}(\tau)H_{b_1}^{\sigma_1}(0)\rangle=\nonumber\\
\frac{1}{Z}\sum_k\sum_{n,m=0}^\infty
\frac{(\tau-\beta)^n(-\tau^m)}{n!m!}\langle\psi_k|H^nH_{b_2}^{\sigma_2}H^m
H_{b_1}^{\sigma_1}|\psi_k\rangle
\label{eq:product}
\ee 
where $\tau$ is the imaginary time, $Z$ is partition function
and the summation over $n$ and $m$ 
comes from Taylor-expansion of $e^{(-\beta+\tau)H}$ and $e^{-\tau H}$. 
Following Ref.\onlinecite{sandvik2} the relation (\ref{eq:product}) 
can be simplified to
\be
\Bigl\langle \sum_{m=0}^{n_s-2}
\frac{(\beta-\tau)^{n_s-m-2}\tau^m}{\beta^{n_s}}
\frac{(n_s-1)!}{(n_s-m-2)!m!}N_m^{b_1b_2,\sigma_1\sigma_2}\Bigr\rangle_W
\label{eq:SSE_current}
\ee
where $n_s$ is the length of sequence of the local operators and it changes 
in each QMC sampling. $N_m^{b_1b_2,\sigma_1\sigma_2}$ is the number of times 
that two operators $H_{b_1}^{\sigma_1}$
and $H_{b_2}^{\sigma_2}$ appear in this sequence with distance 
of $m$ local operators, and $\langle...\rangle_W$ indicates an
arithmetic average using configurations with relative weight $W$.
In this work we introduce an efficient way to sample by SSE the
current-current response function. To this end, we multiply expression
(\ref{eq:SSE_current}) by $e^{i\omega_n\tau}$ and integrate over the
imaginary time $\tau$, we obtain:
\be
\frac{1}{\beta}\Bigl\langle\sum_{m=0}^{n_s-2} ~_1F_1(m+1,n_s;2i\pi
n)N_m^{b_1b_2\sigma_1\sigma_2}\Bigr\rangle_W
\ee
where 
\be
~_1F_1(m+1,n;z)=\nonumber\\
\frac{(n-1)!}{(n-m-2)!m!}\int_{0}^{1}dx \exp(zx)
x^{m}(1-x)^{n-m-2}
\ee
is the confluent hypergeometric function.

Therefore, the  current-current correlation 
acquires contributions determined by length of operator string $n_s$.
All these contributions are stochastically sampled in an 
efficient way, and in  each statistical measurement 
the correlation function $\Lambda(q=0, i\omega_n)$ 
has the following estimator:
\be
\frac{-1}{\beta}\sum_{\sigma_1,\sigma_2=\pm}\sigma_1\sigma_2\sum_{m=0}^{n_s-2} 
{_1F}_1(m+1,n_s;2i\pi n)N_m^{\sigma_1\sigma_2}
\ee
where 
$N_m^{\sigma_1\sigma_2}=\sum_{b_1,b_2}N_m^{b_1b_2,\sigma_1\sigma_2}$.

\begin{figure}[t]
\includegraphics[width=.8\hsize]{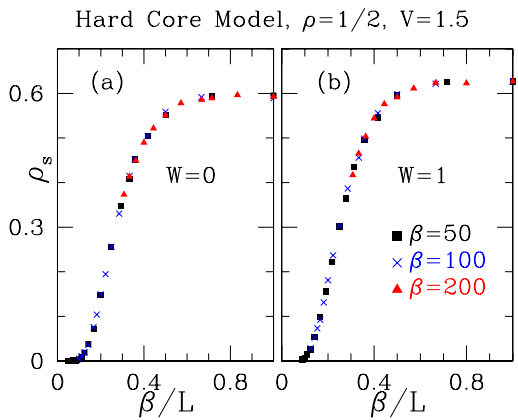}
\caption{(color online) Superfluid stiffness for an integrable (a) and
a non-integrable (b) model versus $\beta/L$. The system size $L$ is ranging 
from 50 to 1200.
}
\label{rhosV1.5L}
\end{figure}

\begin{figure}[t]
\includegraphics[width=.8\hsize]{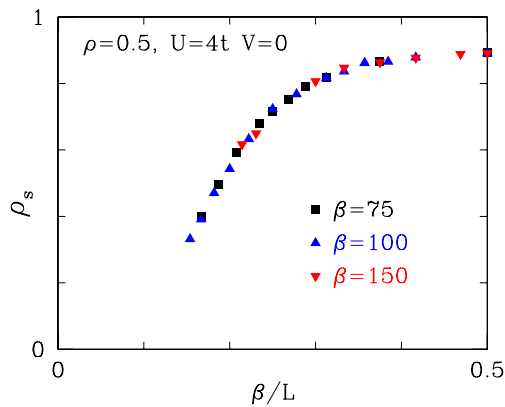}
\caption{(color online) Superfluidity of the soft-core bosons versus
scaled system size at half filling, the on-site interaction is $U=4$}
\label{rhosU2rH}
\end{figure}

{\it Discussion:}
At zero temperature,  for non degenerate ground state,  the Drude weight and
the superfluidity are the same. 
In a 1D system at any finite temperature $\rho_s$ is expected to be 
zero in the
thermodynamic limit, whereas  the Drude weight can be non-zero. 
For hardcore and softcore bosons in a 1D lattice, a systematic size
scaling of the superfluid density $\rho_s$ 
clearly shows that this quantity  vanishes  in the thermodynamic limit
and for  any finite temperature (see figures \ref{rhosV1.5L} 
and \ref{rhosU2rH}). 
Further, we find that, for a fixed set of
parameters and at {\em half filling}, 
all superfluidity data versus $1/L$ collapse to one curve
whenever  the $x$-axis is appropriately scaled with the 
temperature $T$ (see
figures \ref{rhosV1.5L} and \ref{rhosU2rH}).  This analysis suggests the 
scaling form  $\rho_s(\beta, L)\equiv\rho_s(\beta/L)$. 
If one takes the order of limit $T\rightarrow0$ after
$L\rightarrow\infty$, superfluidity remains zero even at zero
temperature. Notice that 
 by taking first the limit  $T\rightarrow0$ and then
$L\rightarrow\infty$ superfluidity has a finite value for  the gapless
phase, but this is not a  signature of superfluidity,  rather the 
occurrence of a finite zero temperature Drude weight.
Though in 1D is not possible to have a finite superfluid density 
at any non zero temperature, several authors have identified  
the  finite zero temperature Drude weight with the superfluid density 
for a superfluid with vanishing critical temperature. 
We believe that this identification is a bit confusing and therefore we
prefer to think about absence of superfluidity and superconductivity in
1D systems, as commonly reported in the textbooks.

\begin{figure}[t]
\includegraphics[width=.8\hsize]{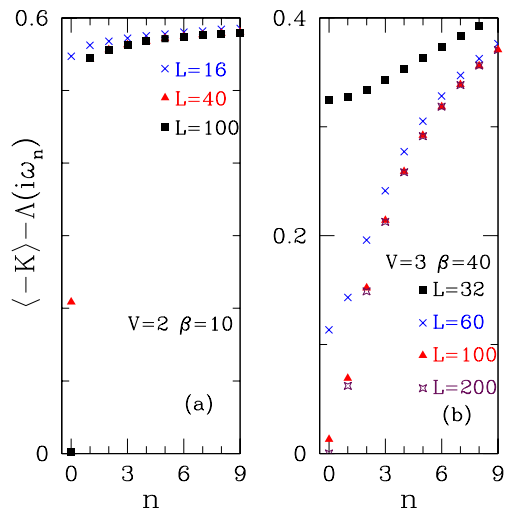}
\caption{(color online) $(a)$ Current-current correlation for an integrable
model in the metallic phase. The zero frequency data shows superfluidity
while the extrapolation to $n\rightarrow0$ is the Drude weight. $D$ remains
finite with increasing $L$ while $\rho_s$ vanishes. 
$(b)$ In the insulating phase $D$ and $\rho_s$ have the same value 
and both tend to zero by increasing $L$.}
\label{DV2W0}
\end{figure}

Fig.~\ref{DV2W0} shows the current-current correlation versus $\omega_n$ in
the metallic and insulating phases of an integrable model ($W=0,~U=\infty$). 
The zero-frequency value is the superfluid density $\rho_s$  and the limit
$\omega_n\rightarrow0$ gives the  Drude weight $D$. For  $W=0$ 
at zero temperature, there exists a critical value
 $V_c/t=2$ below which the Drude weight is finite. In the
first case ($a$) shown in Fig.(\ref{DV2W0}) 
 with $V/t=2$ the  Drude weight has a finite value at any finite
temperature,  which is consistent with the previous works\cite{zotos}. In the
insulating phase (case $b$) with $V/t=3$, the superfluid density 
 coincides with the Drude
weight and they both tend to zero as the system size increases.

%\begin{figure}[t]
%\includegraphics[width=.8\hsize]{rhosV1.5W1L.eps}
%\caption{(color online) Superfluidity for the Hord core bosons with
%nearest neighbor (n.n) and the next n.n interactions, versus $\beta/L$ at
%half filling.
%}
%\label{rhosV1.5W1}
%\end{figure}

%Two lines shown in the plot are
%extrapolations for $L=200$ and $L=300$. The inset shows size dependence of 
%$\rho_s$ and $D$, with increasing $L$ they decrease, $D$ reaches a saturated
%value while $\rho_s$ tends to zero, in fact the data indicates
%nonmonotonic behavior of Drude weight. 
%This nonmonotonic size dependency becomes even more significant at higher
%temperatures, this might suggest the contrary results obtained by
%reference \onlinecite{zotos} with ours for a non-integrable model. 
%To understand the behavior of Drude weight 
%in these models one has to go large system sizes
%such that finite size effect totally disappear. 

\begin{figure}[t]
\includegraphics[width=.8\hsize]{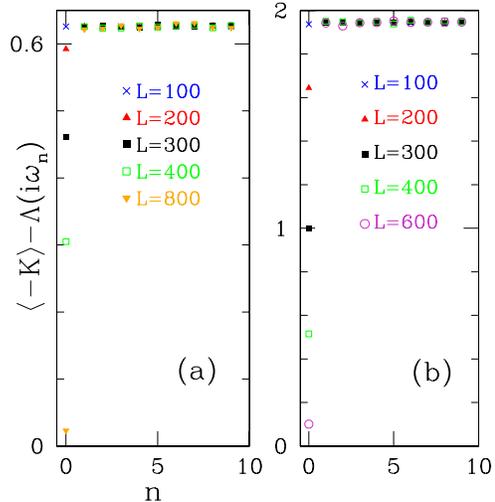}
\caption{(color online) Response function vs. $n$ for (a) hardcore bosons
with $V/t=1.5$, $W/t=1$, $T/t=1/100$ and (b) Bose-Hubbard model with softcore
constraint and  $U/t=2$,
$\mu/t=-0.4$, $T/t=1/25$. The system sizes ranges from $L=100$
to $L=800$.}
\label{DV1.5W1}
\end{figure}

In a non-integrable model such as hard-core bosons with nearest and
next nearest neighborer interactions earlier works have suggested zero 
Drude weight as
system size increases. With SSE we can go to very large system sizes
and low temperatures  and check the scaling dependence of the Drude
weight. In Fig.~\ref{DV1.5W1} we have plotted
current-current correlation versus Matsubara frequency for different
$L$, and a fixed temperature $T=1/100$. 
As shown in the same Figure (\ref{DV1.5W1}) 
we have also found a finite Drude weight at finite $T$ in the celebrated 
Bose-Hubbard model with  softcore constraint and in several other models 
(not shown).  
Although some evidence that few particular  non integrable models 
could have a finite Drude weight at finite temperature have been reported  
before, here we have 
found a very convincing evidence that this behavior should be 
generic for 1D gapless system regardless from their integrability. 
We have supported this statement by state of the art numerical 
 calculations obtained for very large system sizes and low temperature so that 
all possible extrapolations are perfectly under control.   

In conclusion it turns out that, at low energy,  all gapless lattice 
models studied  scale  to the Luttinger liquid fixed point 
where the backscattering is a marginally 
irrelevant coupling and the current is therefore  conserved at the fixed point.
This is therefore a peculiar and generic feature of 1D. 
Indeed  in 2D systems, such as hardcore bosons with n.n. 
repulsion in a square and triangular
lattice, we found no difference between $\rho_s$ and $D$. 

%\acknowledgements
\acknowledgments
We thank M. Troyer for useful discussions. This work is partially 
supported by COFIN-2005 and CNR.

%%%%%%%%%%%%%%%%%%%%%%%%%%%%%%%%%%%%%%%%%%%
%% The following lines show an example how to produce a bibliography
%% without the help of the BibTeX program. This could be used instead
%% of the above.
%%%%%%%%%%%%%%%%%%%%%%%%%%%%%%%%%%%%%%%%%%%


\begin{thebibliography}{9}
\bibitem{ceperley}
E. L. Pollock and D. M. Ceperley Phys. Rev. B {\bf 36}, 8343 (1987).

\bibitem{scalettar}  G. G. Batrouni, R. T. Scalettar and G. T. Zimanyi 
Phys. Rev. Lett. {\bf 65}, 1765 (1990), ibidem Phys. Rev. B {\bf 46}, 
9051 (1992).

\bibitem{dmrgbose} L. I. Plimak, M. K. Olsen, and M. Fleischhlauer 
Phys. Rev. A {\bf 70}, 013611 (2004).

\bibitem{troyer} S. Wessel, F. Alet, M. Troyer, and G.
G. Batrouni, Phys. Rev. A {\bf 70}, 053615 (2004).

\bibitem{sandvik}  A. W. Sandvik, Phys. Rev. B {\bf 56}, 11678 (1997).

\bibitem{fisher}
M. P. A. Fisher, P.B. Weichman, G. Grinstein and D. S. Fisher,
Phys. Rev. B {\bf 40}, 546 (1989).

\bibitem{cazalilla} see e.g. 
M. A. Cazalilla J. Phys. B {\bf 37}, S1 (2004) and references therein.
 

\bibitem{greiner2d} M. Greiner {\em et al.} Nature (London) {\bf 415}, 39 (2002). 

\bibitem{greiner3d} M. Greiner {\em et al.} Nature (London) {\bf 426}, 537 (2003). 

\bibitem{boh} see e.g. 
 H. Moritz {\it et al.} \prl {\bf 94}, 210401 (2005) 
and references therein.

\bibitem{andrei} A. Rosch and N. Andrei Phys. Rev. Lett. {\bf 85}, 1092 (2000).

\bibitem{zotos}
X. Zotos and P. Prelov\"sek Phys. Rev. B {\bf 53}, 983 (1996). 

\bibitem{zotos2}
H. Castella, X. Zotos and P. Prelov\"sek Phys. Rev. Lett. {\bf 74}, 972
(1995). 

\bibitem{poilblanc} D. Poilblanc and {\it et al.}, Europhys. Lett. {\bf
22}, 537 (1993).

\bibitem{hanke}
S. Kirchner, H. G. Evertz and W. Hanke  Phys. Rev. B {\bf 59}, 1825
(1999).

\bibitem{kawakami}  S. Fujimoto and N. Kawakami, Phys. Rev. Lett. 
{\bf 90}, 197202 (2003); {\it ibid} S. Fujimoto and N. Kawakami 
Jour. Phys. A {\bf 31}, 465 (1998).

\bibitem{sandvik2}  A. W. Sandvik, J. Phys. A {\bf 25}, 3667 (1992).

\bibitem{syljuasen} O. F. Syljuasen and A. W. Sandvik, Phys. Rev. E {\bf
66}, 046701 (2002).

\bibitem{scalapino}  D. J. Scalapino, S. R. White and S. Zhang 
Phys. Rev. B {\bf 47}, 7995 (1993).

\bibitem{note} In principle there is a subtle issue related to the 
$\omega \to 0$ limit, that should be employed for real frequencies. 
We assume here that the analytic continuation of the function $\Lambda (i \omega_n)$ is possible, as it is obvious on any finite cluster, and therefore 
this limit can be obtained by interpolation of Matsubara frequencies 
around $\omega=0$, namely at small enough temperatures. 


\end{thebibliography}
\end{document}